\documentclass[aps, twocolumn, 8pt]{revtex4}
\usepackage{latexsym}
\usepackage{graphicx}

\usepackage{amssymb}
\def\II{\mathbb I}

\newtheorem{definition}{Definition}
\newtheorem{lemma}{Lemma}
\newtheorem{theorem}{Theorem}
\newtheorem{corollary}{Corollary}

\begin{document}

\title{Group Covariant Protocols for Quantum String Commitment}
\author{Toyohiro Tsurumaru}
\affiliation{Mitsubishi Electric Corporation,\\
Information Technology R\&D Center\\
5-1-1 Ofuna, Kamakura-shi, Kanagawa,
247-8501, Japan}

\begin{abstract}
We study the security of quantum string commitment (QSC) protocols with group covariant encoding scheme.
First we consider a class of QSC protocol,
which is general enough to incorporate all the QSC protocols given in the preceding literatures.
Then among those protocols,
we consider group covariant protocols and show that
the exact upperbound on the binding condition can be calculated.
Next using this result,
we prove that for every irreducible representation of a finite group,
there always exists a corresponding nontrivial QSC protocol
which reaches a level of security impossible to achieve classically.
\vskip10pt
PACS number(s): 03.65.-a, 03.67.Dd, 89.70.+c
\end{abstract}

\maketitle

\section{Introduction}
Commitment is an important building block of classical cryptographic protocols.
Informally, commitment protocols in general provide the function of a safe or envelope
that can be exchanged over a communication channel;
first the sender Alice sends an evidence of data $x$ of her choice
to the receiver Bob without revealing $x$ itself.
After some time Alice will reveal $x$,
and then Bob can verify that it is indeed the original value of $x$ that she chose
by inspecting the evidence received before.

With the help of computational intractability assumptions,
such task can easily be realized,
in such a way that the secrecy of $x$ against Bob and the unchangeability (the binding condition) of
$x$ by Alice are both perfectly fulfilled\cite{Goldreich}.
However, when it comes to the construction of unconditionally secure protocols,
things change drastically.
It was proved by Lo and Chau\cite{Lo}, and also by Mayers independently\cite{Mayers},
that such a protocol with perfect secrecy and binding,
or the so-called bit commitment (BC), is in fact impossible even by quantum protocols.

Among many attempts to circumvent this no-go theorem,
we focus here on {\it quantum string commitment},
or QSC for short\cite{Kent, Tsurumaru, Buhrman2}.
In QSC protocols,
the sender is supposed to commit $n>1$ bits of data in a single session of protocol,
and we are no more interested in fulfilling both the secrecy and the binding conditions perfectly.
Instead we study a trade-off between the two conditions.
In general, partial information about $x$, say $b$ bits, may be accessible to Bob prior to the reveal phase,
and on the contrary, Alice may be able to change $a$ bits after the commitment phase.
Still, as long as $a+b<n$,
such a scheme provides a nontrivial quantum cryptographic protocol
in that it reaches a classically impossible level of security.
Indeed a number of protocols have been obtained
that are nontrivial in this sense\cite{Kent, Tsurumaru, Buhrman2}.

In this paper,
we consider QSC protocols which have group covariant commitment state $\rho_x$'s
and study its security in terms of the security criteria given by Buhrman et al.\cite{Buhrman2}.
First we consider a class of QSC protocols,
which is general enough to incorporate all the QSC protocols defined explicitly in the preceding literatures.
Then we show that if the encoding scheme for such protocol is covariant under an irreducible representation of a group $G$,
one can calculate the exact upper bound on its binding condition.
Next combining this result with the well-known theorems for quantum optimum detection problem
with covariant input states,
we prove that for every irreducible representation of a finite group $G$,
there always exists a nontrivial QSC protocol.
In other words, we demonstrate how to construct infinitely many types of nontrivial QSC protocols
with $a+b<n$.

\section{Quantum String Commitment}
\subsection{Description of Protocol}
A quantum string commitment (QSC) protocol is a quantum communication
protocol between two parties, the sender Alice and the receiver Bob,
which consists of two stages, the commit phase and the reveal phase.
\begin{itemize}
\item ({\it Commit Phase}) If both parties are honest, Alice chooses a string $x\in\{0,1\}^n$.
From Bob's point of view, string $x$ has probability $p_x$.
Alice and Bob communicate.
Let $\rho_x$ denote Bob's state at the end of the protocol if Alice committed string $x$.
\item ({\it Reveal Phase}) If both parties are honest, Alice sends $x$ and other reveal information to Bob.
Bob accepts.
\end{itemize}
In addition, for the sake of simplicity,
we assume that honest Alice chooses $x\in\{0,1\}^n$ with a uniform distribution $p_x=2^{-n}$.

\subsection{Security Requirements}
As was the case for bit commitment,
there are two conditions of security for quantum string commitment,
that is, the {\it secrecy} condition and the {\it binding} condition.
While there are various ways of defining them\cite{Kent, Tsurumaru, Buhrman2}, especially for binding,
in this paper we use the most simple of them given in Ref.\cite{Buhrman2},
based on accessible information $I_{\rm acc}$.

The concealing condition, or the secrecy,
deals with cases where Alice is honest.
Malicious Bob in general does anything possible to obtain information regarding $x$ prior to the reveal phase,
and in order to discuss the security there, we want to bound the amount of his information from above.
The relevant quantity for such purpose is the accessible information $I_{\rm acc}$ for
the ensemble of commitment states ${\cal E}=\{p_x,\rho_x\}$.
\begin{definition}[Concealing Condition]
A QSC protocol is $b$-concealing if $I_{\rm acc}({\cal E})\le b$.
Here $I_{\rm acc}({\cal E})$ is Bob's accessible information measured at the end of the commit phase.
\end{definition}
As pointed out by Buhrman et al.\cite{Buhrman2},
the stronger notion of Holevo Information $\chi$ is not appropriate for this purpose
since in many cases $\chi$ overestimates $I_{\rm acc}$ and can set $b$ larger than the reality.

On the other hand,
the binding condition applies when Bob is honest.
It is possible that malicious Alice may postpone her decision on the value of $x$
until after the commit phase,
and try to reveal one of several different values of $x$ at the reveal phase.
In order to limit Alice's attack of this type,
we employ the following security criterion.
\begin{definition}[Binding Condition]
A QSC protocol is $a$-binding if
$\sum_{x\in\{0,1\}^n}\tilde{p}_x\le2^a$, where $\tilde{p}_x$
is the probability that Alice is able to successfully reveal $x\in\{0,1\}^n$
at the reveal phase.
\end{definition}
For purely classical protocol without any special assumption,
such as computational intractability or relativistic constraints,
$a+b\ge n$ always holds.
This can be shown in a similar way to the proof of the impossibility
of information-theoretically secure bit commitment\footnote{%
See e.g., Ref.\cite{Goldreich}, Chapter 4, Exercise 32.}.
Hence, as long as $a+b<n$ is satisfied,
we consider a quantum protocol to be nontrivial.


\section{Group Covariant Protocol}
\subsection{Basic Scheme}
\subsubsection{Description}
From now on,
we restrict ourselves to the following type of QSC protocols.
This scheme allows us to convert an arbitrary ensemble of states ${\cal E}=\{p_x,\rho_x\}$
to a corresponding QSC protocol in a straightforward way.
Moreover, as will be shown below,
it is general enough to incorporate all previous QSC protocols appearing
in preceding literatures\cite{Kent, Tsurumaru, Buhrman2} without sacrificing security.
\begin{itemize}
\item ({\it Commit Phase})
Honest Alice generates a state vector $|\psi_x\rangle\in H_A\otimes H_B$,
which depends on the value of $x$ she chooses,
and sends its second half (in $H_B$) to Bob.
\item ({\it Reveal Phase})
Alice sends to Bob the remaining half of her state.
Honest Bob measures it projectively with respect to $|\psi_x\rangle$,
and outputs ACCEPT if and only if the outcome is correct.
\end{itemize}
Bob's view at the end of the commit phase
is of course $\rho_x={\rm Tr}_A|\psi_x\rangle\langle \psi_x|$.
Thus according to Definion 1,
the secrecy is measured by the accessible information $I_{\rm acc}({\cal E})$ for the ensemble
${\cal E}:=\{p_x,\rho_x\}$.

\subsubsection{Binding Condition}
For the above scheme,
Alice's cheating strategy can always be formulated as follows.
As in the proof of the no-go theorem of quantum BC\cite{Lo,Mayers},
it is convenient to adopt the {\it decoherence} point of view
by introducing a suitable environment Hilbert space.
Then without loss of generality,
we may assume that the state shared between two parties at the end of the commit phase
is a pure state $|\Psi\rangle\in H_{\tilde{A}}\otimes H_B$.
Here the dimension of $H_{\tilde{A}}$ is assumed to be arbitrary, say $d_{\tilde{A}}$.
Subsequently in reveal phase,
Alice performs generalized quantum operations\cite{Nielsen} on $H_{\tilde{A}}$,
\begin{eqnarray}
\lefteqn{O_x:=\{E_{xi}\ |\ i=1,\dots,m\},}\hspace{2.5cm}\label{eq:def_M_xi}\\
\sum_{i=1}^m E_{xi}^\dagger E_{xi}&=&\II_{d_{\tilde{A}}},\nonumber
\end{eqnarray}
which depend on the value of $x$ that she wishes to reveal,
and sends the obtained quantum state to Bob.
Quantum operation $O_x$ yields classical outcome $i$ with probability
\[q_{xi}:={\rm Tr}_{{\tilde{A}}B}\left[E_{xi}|\Psi\rangle\langle\Psi|E_{xi}^\dagger\right],\]
as a result of which Bob obtains $E_{xi}|\Psi\rangle\langle\Psi|E_{xi}^\dagger/q_{xi}$.
Bob then measures it projectively with respect to $|\psi_x\rangle$,
and accepts $x$ with probability $\tilde{p}_x=\sum_i\left|\langle\psi_x|E_{xi}|\Psi\rangle\right|^2$.
Hence the binding condition is measured by
\begin{equation}
\sum_x\tilde{p}_x\le\max_{\Psi}\sum_x\max_{O_x}\sum_i\left|\langle\psi_x|E_{xi}|\Psi\rangle\right|^2.
\label{eq:tilde_p_x}
\end{equation}

\subsubsection{Relation to The Existing Protocols.}
Here we show that all previous QSC protocols appearing
in preceding literatures\cite{Kent, Tsurumaru, Buhrman2} can be converted to
our basic scheme without sacrificing security.

This is trivial for those protocols defined in Ref.\cite{Kent,Tsurumaru},
where honest Alice sends to Bob
a pure state which is not entangled with any of her state.
In this case $H_A$ is considered as a one-dimensional vector space.

The conversion is also possible for $LOCKCOM$-type QSC protocols\cite{Buhrman2},
where the sender is supposed to choose
random number $i\in\{1,\dots,R\}$ besides $x$, and send $U_i|x\rangle$ in commit phase,
with $U_i$ being a unitary operator.
For such protocols,
one simply needs to choose $|\psi_x\rangle$ for the converted protocol as
\[|\psi_x\rangle:=\frac1{\sqrt{R}}\sum_i|i\rangle_A\otimes U_i|x\rangle_B.\]
This is a purification of $\rho_x$ of the original protocol,
i.e.,
$\rho_x = \frac1R\sum_{i=1}^R U_i|x\rangle_{BB}\langle x|U_i^\dagger$
and $\rho_x={\rm Tr}|\psi_x\rangle\langle\psi_x|$.
Clearly, secrecy is not changed with such conversion.
Binding can also be guaranteed due to the following argument;
In the reveal phase of the original protocol,
Bob uses an operator
\[P=\sum_i|i\rangle_{AA}\langle i|\otimes U_i|x\rangle_{BB}\langle x|U_i^\dagger\]
to test the state obtained,
while for the converted version
$\tilde{P}=|\psi_x\rangle\langle\psi_x|$ is used.
$P$ and $\tilde{P}$ are projection operators commuting with each other
and $\tilde{P}$ is of smaller rank.
Thus any strategy by Alice for the converted protocol will always give an equal
or higher success probability when applied to the original protocol.

\subsection{Group Covariant Protocols}
If we restrict ourselves to group covariant protocols,
to be defined shortly,
we can in fact calculate the maximum value of $\sum_x\tilde{p}_x$ exactly.
This is because, as we will show below,
any cheating strategy by malicious Alice is equivalent to choosing $|\Psi\rangle$ of Eqn.(\ref{eq:tilde_p_x})
such that $\rho={\rm Tr}_A|\Psi\rangle\langle\Psi|$ is a group invariant state.
Especially when a protocol is invariant under an irreducible representation of group $G$,
it means that $\rho$ must be proportional to unit vector $\II_d$
and this fact greatly simplifies calculations.

\subsubsection{Irreducible Representation}
As a preliminary to this result,
we introduce some terminology of group theory \cite{Davies}.
Representation $D$ of a group $G$ is a set of matrices
$\{\, D(g)\, |\, g\in G\, \}$,
satisfying $\forall g_1,\forall g_2\in G$, $D(g_1)D(g_2)=D(g_1g_2)$.
In what follows we suppose that $D(g)$'s are $d\times d$ unitary matrices
operating on $d$-dimensional vector space $H_B$.
Representation $D$ is irreducible when no nontrivial vector subspace of $H_B$
is invariant under $G$.
It is a direct consequence of Shur's lemma that for irreducible $D$,
a $d\times d$ matrix $M$ commutes with $D(g)$, $\forall g\in G$
iff $M$ is proportional to the unit matrix $\II_d$.

Bob's view $\{\,\rho_x\,|\,x\in\{0,1\}^n\,\}$,
which we introduced above,
is called {\it covariant} if it is invariant as a set under operations of $G$.
In other words,
\begin{equation}
\forall x,\forall g\in G,\exists y,\ \rho_y = D(g)\rho_x D^\dagger(g).
\label{eq:covariance_of_rho}
\end{equation}
The action of $G$ on a covariant set $\{\rho_x\}$ is called {\it transitive}
if for all $x$ and $y$ there exists $g\in G$ such that $\rho_y =D(g)\rho_x D^{\dagger}(g)$.

In the rest of this paper,
we will refer to a QSC protocol as {\it group covariant protocol}
if it possesses $\rho_x$'s transforming covariantly and transitively
under an irreducible representation of a finite group $G$.

\subsubsection{Symmetrized Strategy}
Using the above notations,
we shall show that any strategy used by a malicious Alice can always be
converted into an equally effective form
in which she commits a symmetric state.

As explained in the paragraph around Eqn.(\ref{eq:def_M_xi}),
Alice's cheating strategy can always be characterized by state
$|\Psi\rangle\in H_{\tilde{A}}\otimes H_B$ that she generates during commit phase
and the set of quantum operations given in Eqn.(\ref{eq:def_M_xi}).
The first key observation is that instead of using $|\Psi\rangle$,
she may as well introduce an ancillary Hilbert space $H_{A'}$ and generate
\[
|\Phi\rangle_{A'\tilde{A}B} = \frac1{\sqrt{|G|}}\sum_{g\in G}|g\rangle_{A'}\otimes D_B(g)|\Psi\rangle_{\tilde{A}B},
\]
with $D_B(g)$ acting on $H_B$.
The set of states $\{|g\rangle_{A'}\}_{g\in G}$ form an orthogonal basis labeled by $G$,
$\langle g|g'\rangle_{A'}=\delta_{g,g'}$.
With such $|\Phi\rangle$,
Alice can achieve a value of $\sum_x\tilde{p}_x$ at least equal to the original attack,
e.g., by first measuring $|g\rangle_{A'}$ in the reveal phase,
and then operating on $H_{\tilde{A}}$ with $O_x$ with a permuted value of $x$.
Note that in this case $D_B(g)$ merely permutes the values of $\tilde{p}_x$ and
the sum of $\tilde{p}_x$ remains unchanged.

On the other hand, $|\Phi\rangle$ as seen from Bob,
or $\sigma:={\rm Tr}_{A',\tilde{A}}|\Phi\rangle\langle\Phi|$,
is clearly invariant under $G$,
meaning that it must be proportional to the unit matrix, $\sigma=\II_d/d$.
Hence Alice's best strategy during commit phase
is to send Bob the maximally entangled state
\[|\Phi_{\rm ME}\rangle:=\frac1{\sqrt{d}}\sum_{a=1}^d|a\rangle_A\otimes|a\rangle_B.\]
Subsequently in reveal phase,
Alice's operations in general can be described, as in the original attack,
by a set of operators $O_x$, as defined in Eqn.(\ref{eq:def_M_xi}),
although the actual form of $O_x$'s achieving the maximum $\sum_x\tilde{p}_x$
may not be the same as those used in the original attack.
Hence without loss of generality,
we may assume $\tilde{p}_x$ takes the form
\[
\tilde{p}_x=\max_{O_x}\sum_{i=1}^m\left|\langle\psi_x|E_{xi}|\Phi_{\rm ME}\rangle\right|^2.
\]
It is easy to see that due to the symmetry properties of our protocol,
the maxima of $\tilde{p}_x$'s are all equal for any value of $x$.
Thus it remains to maximize $\tilde{p}_x$ for an arbitrarily chosen value of $x$, say $\tilde{p}_0$.

\subsubsection{Maximizing $\tilde{p}_0$}
Recall that $|\Phi_{\rm ME}\rangle$ is invariant under $U_A\otimes U_B$
with $U_A$ being an arbitrary unitary transformation and $U_B$ its complex conjugate.
Thus by appropriate choice of orthonormal bases $\{|\mu_a\rangle\}$ and $\{|\nu_a\rangle\}$
and using the Schmidt decomposition,
we can rewrite $|\Phi_{\rm ME}\rangle$ and $|\psi_0\rangle$ as
\begin{eqnarray*}
|\Phi_{\rm ME}\rangle&=&\sum_a\frac1{\sqrt{d}}|\mu_a\rangle_A\otimes|\mu_a\rangle_B,\\
|\psi_0\rangle&=&\sum_a\sqrt{\lambda_a}|\nu_a\rangle_A\otimes|\mu_a\rangle_B,
\end{eqnarray*}
where $\lambda_a$'s are the eigenvalues of $\rho_0$,
and by symmetry, of all $\rho_x$'s.
Then by decomposing $E_{0i}$ as
\[E_{0i}=\sum_{a,b}N^i_{ab}|\nu_a\rangle\langle\mu_a|,\]
$\tilde{p}_0$ can be expressed as
\[\tilde{p}_0=\frac1d\sum_i\left|\sum_a N^i_{aa}\lambda_a^{1/2}\right|^2\]
with
\[\sum_i\sum_a \left(N^i_{ab}\right)^*N^i_{ac}=\delta_{bc}.\]
It is convenient to interpret the diagonal elements of $N^i_{ab}$ as
an $m$-dimensional vector $\vec{v}_a=(N^1_{aa},\dots,N^m_{aa})$.
The lengths of $\vec{v}_a$'s are smaller than one since $|\vec{v}_a|^2\le \sum_i\sum_b|N^i_{ba}|^2=1$.
With this property, $\tilde{p}_0$ can be bounded from above as
\begin{eqnarray*}
\tilde{p}_0&=&\frac1d\left|\sum_a \lambda_a^{1/2} \vec{v}_a\right|^2
\le\frac1d\left|\sum_a \lambda_a^{1/2}\right|^2\max_a\left|\vec{v}_a\right|^2\\
&\le&\frac1d\left|\sum_a \lambda_a^{1/2}\right|^2.
\end{eqnarray*}
with the equality holding for $m=1$ and $N^1_{ab}=\delta_{ab}$.
Summarizing the above, we obtain the following theorem.
\begin{theorem}[Exact Upperbound on Binding]
For a group covariant QSC protocol,
and $\lambda_a$'s being the eigenvalues of $\rho_x$,
\begin{equation}
\sum_x\tilde{p}_x\le\frac{2^n}{d}\left|\sum_a \lambda_a^{1/2}\right|^2
\label{eq:upper_bound_on_tilde_p}
\end{equation}
with the equality holding for Alice's attack using the maximally entangled state.
\end{theorem}
In terms of Renyi entropy $S_\alpha$,
Eqn.(\ref{eq:upper_bound_on_tilde_p}) can be rewritten
in a form similar to Theorem 2 of Ref.\cite{Buhrman2}:
\begin{equation}
\log\left(\sum_x\tilde{p}_x\right)\le n-\left[S(\rho)-S_{1/2}(\rho_0)\right],
\label{eq:upper_bound_on_tilde_p_2}
\end{equation}
where $S_{1/2}(\cdot)$ denotes Renyi entropy for $\alpha=1/2$.
The mixed state $\rho$ is defined as $\rho=\sum_x p_x\rho_x=\II/d$.

\subsection{Example: Tetrahedral Encoding}
As an application of Theorem 1, we consider $\rho_x$'s covariant under the tetrahedral group\cite{Davies}.

Define qubit states
\[
\begin{array}{cc}
|\xi;00\rangle=\left(
\begin{array}{cc}
1\\0
\end{array}
\right),
&
|\xi;01\rangle=\left(
\begin{array}{cc}
\sqrt{1/3}\\ \sqrt{2/3}
\end{array}
\right),
\\
|\xi;10\rangle=\left(
\begin{array}{cc}
\sqrt{1/3}\\ \sqrt{2/3}\,\omega
\end{array}
\right),
&
|\xi;11\rangle=\left(
\begin{array}{cc}
\sqrt{1/3}\\ \sqrt{2/3}\,\omega^2
\end{array}
\right)
\end{array}
\]
with $\omega = e^{2\pi i/3}$.
These four states are covariant under an irreducible representation of
the tetrahedral group $T$,
which we will denote as $D(g)$, for $g\in T$
\footnote{%
To be precise, $|\xi;xy\rangle$'s are covariant under
the subgroup of $SU(2)$ that is homomorphic to $T$.}.
Now assume that $n$ is an even number.
Also define $|\psi_x\rangle\in H_B$ as
\[
|\psi_x\rangle:=|\xi;x_1x_2\rangle\otimes\cdots\otimes|\xi;x_{n-1}x_n\rangle.
\]
and let $H_A$ be a one-dimensional complex vector state.
That is, honest Alice is supposed to send Bob pure state $|\psi_x\rangle$ in commit phase.

Such $\rho_x$'s are covariant under $G:=T\times\cdots\times T$
with its irreducible representation $D(g_1)\otimes\cdots\otimes D(g_{n/2})$.
Thus applying Theorem 1 obtained above,
we readily find the exact upper bound on binding;
$\sum_x\tilde{p}_x\le2^{n/2}$.
In other words, this protocol is $n/2$-binding.

Secrecy can also be calculated exactly.
Alice's commitment is $n/2$ independent draws of an ensemble
${\cal E}=\{p_i=\frac14,|\xi;i\rangle\}$ with $i=1,\dots,4$,
and the accessible information for such case is known to be additive,
$I_{\rm acc}({\cal E}^{\otimes \frac{n}2})=\frac{n}2I_{\rm acc}({\cal E})$ \cite{DiVincenzo1,DiVincenzo2}.
Due to this fact and by using the exact value $I_{\rm acc}({\cal E})=\log\frac43$ for the tetrahedral states\cite{Davies},
we find that this protocol is $\frac{n}2\log\frac43$-binding.

In summary, we have $a=\frac{n}2$ and $b=\frac{n}2\log\frac43$ satisfying $n>a+b$,
which is impossible classically.

\section{Secrecy of Covariant Protocols}
At the end of the previous section,
we studied a QSC protocol transforming covariantly under the tetrahedral group
and it turned out to be nontrivial, that is, a classically impossible protocol.
As we will show below,
in fact this is not a coincidence but rather a consequence of symmetric properties of our protocols.

In this section,
by focusing on the cases where all $\rho_x$'s are pure states,
and with the help of the results obtained in the preceding literatures
on the information-theoretic optimum detection problem with covariant input states\cite{Davies},
we will show the following theorem.
\begin{theorem}
For covariant protocols with pure $\rho_x$'s,
either of the following cases holds:
\begin{enumerate}
\item The protocol is equivalent to a purely classical protocols,
i.e.,
all transactions occurring between Alice and Bob are done in computational basis.
\item The protocol is nontrivial, i.e., it satisfies $a+b<n$ with strict inequality.
\end{enumerate}
\end{theorem}

\noindent{\it Proof of Theorem.}
As for secrecy,
there are useful formula giving classical mutual information $I_{\rm acc}$ in a very simple form.
The most relevant among them for our purpose is Lemma 6 of Ref.\cite{Davies},
which reads in our notation as follows.
\begin{lemma}
For a covariant encoding scheme,
the maximum value of accessible information $I_{\rm acc}$ is given by
\begin{equation}
I_{\rm acc}=\log d+\frac{d}{|G|}\sum_{g\in G}\langle\varphi|\rho_g|\varphi\rangle\log\langle\varphi|\rho_g|\varphi\rangle,
\label{eq:theorem_davies}
\end{equation}
where $|\varphi\rangle$ is an appropriately chosen state vector.
\end{lemma}
Mixed state $\rho_g$ appearing in (\ref{eq:theorem_davies}) is indexed by a group element $g\in G$
and is defined as $\rho_g:=D(g)\rho_0D^\dagger(g)$,
where $\rho_0$ denotes $\rho_x$ with $x=0$.
According to Eqn.(\ref{eq:covariance_of_rho}),
every $\rho_g$ equals some $\rho_x$ but the correspondence is not necessarily one-to-one.

Now note that for pure $\rho_x$, the log on the RHS of Inequality (\ref{eq:upper_bound_on_tilde_p}) equals $n-\log d$.
On the contrary, the second term on the RHS of (\ref{eq:theorem_davies}) is clearly no more than zero,
and so, as long as there is at least one nonzero element in the sum of (\ref{eq:theorem_davies}),
the protocol is nontrivial.
Thus it remains to show that a trivial case is always equivalent to a classical protocol.

Clearly, with $\rho_g$ being a pure state,
$\langle\varphi|\rho_g|\varphi\rangle$ can be rewritten as
$\langle\varphi|\rho_g|\varphi\rangle=|\langle\varphi|D(g)|\psi_0\rangle|^2$ with $\rho_0=|\psi_0\rangle\langle\psi_0|$.
Then if we suppose that the sum of (\ref{eq:theorem_davies}) is strictly zero,
$|\langle\varphi|D(g)|\psi\rangle|^2=0$ or 1 should hold for all $g\in G$.
Since this quantity should be nonzero at least for one group element $g\in G$,
without loss of generality we may assume $|\varphi\rangle=|\psi_0\rangle$.
Hence we have $\forall g\in G$, $|\langle\psi_0|D(g)|\psi_0\rangle|^2=0$ or 1.
This means that $|\psi_x\rangle$ defined by $\rho_x=|\psi_x\rangle\langle\psi_x|$
are all orthogonal to each other since any $|\psi_x\rangle$ can be
described as $D(g)|\psi_0\rangle$ for some $g\in G$ due to irreduciblilty of $D$.
This completes the proof.

By choosing an arbitrary irreducible representation $D$ of a group $G$,
and with an arbitrary choice of a pure state vector $|\psi\rangle$,
we can always construct a QSC protocol that uses $D(g)|\psi\rangle$ as commitment states.
Moreover, it is clear that for any choice of $G$ and $D$,
there always exists $|\psi\rangle$, such that $D(g)|\psi\rangle$'s
do not form a orthonormal basis,
in which case the obtained QSC protocol is nontrivial due to this Theorem.
Thus we also have the following corollary.
\begin{corollary}
For any irreducible representation $D$ of any finite group $G$,
there always exists a nontrivial QSC protocol with $a+b<n$.
\end{corollary}

\section{Summary}
In this paper,
we introduced a class of QSC protocols and studied its security in terms of
the security criteria given by Buhrman et al.\cite{Buhrman2}.
In particular,
we considered group covariant protocols
and showed how to calculate the exact upper bound on its binding conditions.
Then combining this result with the previously known theorems for the quantum optimum detection problem,
we proved that for every irreducible representation of a finite group $G$,
there always exists a nontrivial QSC protocol.
In other words, we demonstrated how to construct infinitely many types of nontrivial QSC protocols
with $a+b<n$.

A question that arises naturally is for what types of groups and for which representations
we obtain efficient protocols with strong enough security.
In particular, in view of cryptographic applications, such as zero-knowledge proof or message authentication,
$a/n$ and $b/n$ should be minimized.
Although Buhrman et al. have given a protocol that accomplishes arbitrarily small $a/n$ and $b/n$,
their protocol is not efficient.
On the contrary for group covariant schemes as given here,
the obtained protocols are most likely efficient.
Hence it is interesting to investigate our result for other explicit examples of finite groups.

\end{document}